\documentclass[preprint,showpacs]{revtex4}
\usepackage[T2A]{fontenc}
\usepackage[cp1251]{inputenc}
\usepackage[english]{babel}
\usepackage{amsmath}
\usepackage{amsfonts}
\usepackage{graphicx,epsfig}
\usepackage{subfigure}
\usepackage{amssymb}
\usepackage[indent,small]{caption2}
\begin{document}

\title{Optical Bloch oscillation and Zener tunneling in an array of cylindrical waveguides.
Numerical simulation.}

\author{Polishchuk I. Ya.$^{1,2}$, Gozman M. I.$^1$, Polishchuk Yu. I.$^2$}

\affiliation{$^1$ RRC Kurchatov Institute, Kurchatov Sq., 1,
123182 Moscow, Russia\\ $^2$ Moscow Institute of Physics and
Technology, 141700, 9, Institutskii per., Dolgoprudny, Moscow
Region, Russia}

\begin{abstract}
We investigate optical Bloch oscillation, Zener tunneling and
breathing modes in arrays of optical waveguides. We perform a new
method of calculation based on the multiple scattering formalism.
To demonstrate Bloch oscillation and breathing modes, we consider
a planar array of parallel cylindrical waveguides with the
refractive index gradually varying across the array. We
demonstrate that the form of Bloch oscillation may be predicted by
means of dispersion law analysis. To demonstrate Zener tunneling,
we consider a planar array of cylindrical waveguides of two types
situated by turn. The band structure of this array contains two
bands separated by a narrow gap. If the refractive indices of
waveguides gradually vary across the array, the Zener tunneling
leads to the Bloch-Zener oscillation.
\end{abstract}

\maketitle

%%%%%%%%%%%%%%%%%%%%%%%%%%%%%%%%%%%%%%%%%%%%%%%%%%%%%%%%%%%%%%%%%%%%%%

\section{Introduction}

Nowadays, much attention is devoted to arrays of evanescently
coupled optical waveguides which are both of the fundamental and
practical interest. These arrays are useful in integrated optical
circuits and other micro- and nanooptical devices, such as optical
filters and near-field microscopes.

The periodic arrays of optical waveguides represent the particular
case of low-dimensional photonic crystal structures. The general
feature of such systems is the existence of photonic band
structure \cite{Joannopoulos} that is analogous to the electron
band structure in solids. Therefore some effects in optical
lattices may be analogous to some phenomena in solids
\cite{Lederer,Longhi}. In this work we consider optical
counterparts of Bloch oscillation and Zener tunneling.

Around 1930’s it was predicted that an electric field applied to a
crystal should induce an oscillatory motion of the electrons,
known as Bloch oscillation \cite{Bloch,Zener}. Besides, in
multiband systems electrons under an external force can
spontaneously transit from one band to another. This effect is
known as Zener tunneling.

Optical excitations in arrays of waveguides can perform a similar
effects, as it was shown in numerous theoretical
\cite{Wang1,Wang2} and experimental works. The usual pattern to
demonstrate the optical Bloch oscillation and Zener tunneling is a
planar array of parallel waveguides with refractive index linearly
varying across the array. To produce the gradual refractive index
alteration, one can use the thermo-optic
\cite{Pertsch1,Pertsch2,Pertsch3} or electro-optic effects
\cite{Pertsch4}. The other pattern is an array of waveguides of
the same refractive indices, but of different thickness
\cite{Morandotti}. Sometimes array of identical gently curved
waveguides is used \cite{Chiodo,Dreisow1,Dreisow2}.

The optical excitation coupled into such array propagates along
the direction of waveguides oscillating in the transverse
direction, so the propagation way of the excitation takes the
sinusoidal form. This phenomena is the optical counterpart of
electronic Bloch oscillation in solids. The effect of optical
Bloch oscillation can be practically used in different optical
devices for light steering.

The optical counterpart of Zener tunneling may take place in
presence of two bands separated by a gap in the band structure of
the array. The superposition of Bloch oscillation and Zener
tunneling causes the splitting of an optical beam into two beams
propagating along different oscillating ways. This effect is known
as Bloch–Zener oscillation \cite{Wang2,Pertsch3,Dreisow1}.

In most of works, for theoretical simulation of Bloch oscillation
and Zener tunneling the following system of equations is used:

\begin{equation}
\left(i\frac{d}{dz}+\beta_j\right)\,a_j(z)+\gamma\,\Bigl(a_{j-1}(z)+a_{j+1}(z)\Bigr)=0.
\label{Tradi}
\end{equation}

Here the waveguides are assumed to be directed along the $z$-axis,
$j$ is the number of a waveguide, $a_j(z)$ is the amplitude of the
optical excitation at the $j$-th waveguide, $\beta_j$ is the
propagation constant of the $j$-th waveguide, $\gamma$ is the
coupling constant. This system of equations is useful for the
waveguides of any form, but the parameters $\gamma$ and $\beta_j$
should be obtained experimentally.

In this paper we use another method of theoretical simulation
based on multiple scattering formalism (MSF)
\cite{Felbacq1,Felbacq2}. This method is convenient for the arrays
of cylindrical waveguides. Its advantage is that the radii and
refractive indices of the waveguides are the only data required
for the calculation, and one has not to obtain any other
parameters from an experiment. Besides, this method allows to
calculate the spatial distribution of electromagnetic field around
waveguides and inside of them with arbitrary accuracy, as opposed
to Eq. (\ref{Tradi}), that allows only to find the intensity of
optical excitation near every waveguide.

The MSF is explained in Sect. II. In Sect. III we calculate the
band structure of a plane array of infinite cylindrical rods. The
obtained dispersion laws are used in Sect. IV for prediction of
Bloch oscillation of optical beam in an array of rods with
refractive index gradually varying across the array. The
prediction is confirmed by the direct numerical simulation
represented in Sect. V. Besides, in Sect. V the so-called
breathing mode is investigated. In Sect. VI we investigate
Bloch-Zener oscillation of optical beam in a plane array of rods
of two types situated by turns, with gradually varying refractive
indices. Finely, in Conclusion we discuss possible practical
applications of the investigated optical effects and the
possibility of further development of method used in this paper.

%%===============================================================

\section{Multiple scattering formalism}

We consider an array of $N$ parallel dielectric waveguides
directed along the $z$-axis. We assume the waveguides being
infinite cylindrical rods. The array is illuminated by a
monochromatic wave of frequency $\omega$. The velocity of light in
free space is supposed to be unit.

The general idea of multiple scattering formalism is that near the
$j$-th waveguide the incident wave can be represented as a linear
combination of harmonics with certain values of angular momentum
$m$ and longitudinal wave vector $K$:

\begin{equation}
\begin{array}{l} \displaystyle
\mathbf{E}_{inc}(t,\mathbf{r})= e^{-i\omega t}\int dK e^{iKz}
\sum\limits_{m=-\infty}^{+\infty} e^{im\phi^{(j)}}
\Bigl(p_{jm}(K)\,\mathbf{M}^{(1)}_{\omega K m}(r^{(j)})-
q_{jm}(K)\,\mathbf{N}^{(1)}_{\omega K m}(r^{(j)})\Bigr),
\medskip \\ \displaystyle
\mathbf{H}_{inc}(t,\mathbf{r})= e^{-i\omega t}\int dK e^{iKz}
\sum\limits_{m=-\infty}^{+\infty} e^{im\phi^{(j)}}
\Bigl(p_{jm}(K)\,\mathbf{N}^{(1)}_{\omega K m}(r^{(j)})+
q_{jm}(K)\,\mathbf{M}^{(1)}_{\omega K m}(r^{(j)})\Bigr).
\end{array}
\label{EHinc}
\end{equation}

Here $r^{(j)}$ and $\phi^{(j)}$ are the polar coordinates of
two-dimensional vector $\textbf{r}^{(j)}=\{x-x_j,y-y_j\}$, and
$x_j,y_j$ are the coordinates of the axis of the $j$-th waveguide.
Coefficients $p_{jm}(K)$, $q_{jm}(K)$ are called the partial
amplitudes of the incident wave. The expressions for functions
$\mathbf{M}^{(1)}_{\omega K m}(r)$, $\mathbf{N}^{(1)}_{\omega K
m}(r)$ are given in Appendix. One can see that
$\Bigl(\mathbf{N}^{(1)}_{\omega K m}\Bigr)_z(r)=0$, therefore
harmonics containing functions $\mathbf{M}^{(1)}_{\omega K m}(r)$,
$\mathbf{N}^{(1)}_{\omega K m}(r)$ can be named TM- and
TE-harmonics correspondingly.

The wave scattered by the $j$-th waveguide can be represented in
the similar way, but the other functions $\mathbf{M}^{(2)}_{\omega
K m}(r)$, $\mathbf{N}^{(2)}_{\omega K m}(r)$ enter into the
expressions instead of the functions $\mathbf{M}^{(1)}_{\omega K
m}(r)$, $\mathbf{N}^{(1)}_{\omega K m}(r)$:

\begin{equation}
\begin{array}{l} \displaystyle
\mathbf{E}^{(j)}_{sca}(t,\mathbf{r})= e^{-i\omega t}\int dK
e^{iKz} \sum\limits_{m=-\infty}^{+\infty} e^{im\phi^{(j)}}
\Bigl(a_{jm}(K)\,\mathbf{M}^{(2)}_{\omega K m}(r^{(j)})-
b_{jm}(K)\,\mathbf{N}^{(2)}_{\omega K m}(r^{(j)})\Bigr),
\medskip \\ \displaystyle
\mathbf{H}^{(j)}_{sca}(t,\mathbf{r})= e^{-i\omega t}\int dK
e^{iKz} \sum\limits_{m=-\infty}^{+\infty} e^{im\phi^{(j)}}
\Bigl(a_{jm}(K)\,\mathbf{N}^{(2)}_{\omega K m}(r^{(j)})+
b_{jm}(K)\,\mathbf{M}^{(2)}_{\omega K m}(r^{(j)})\Bigr).
\end{array}
\label{EHsca}
\end{equation}

The coefficients $a_{jm}(K)$, $b_{jm}(K)$ are named the partial
amplitudes of the scattered wave. The functions
$\mathbf{M}^{(2)}_{\omega K m}(r)$, $\mathbf{N}^{(2)}_{\omega K
m}(r)$ are also given in Appendix.

The partial amplitudes of incident and scattered waves satisfy to
the following system of equations:

\begin{equation}
\Bigl(S_{jm}(\omega,K)\Bigr)^{-1} \left(\begin{matrix} a_{jm}(K)
\\ b_{jm}(K)
\end{matrix}\right)-
\underset{(l\neq j)}{\sum\limits_{l=1}^N}
\sum\limits_{n=-\infty}^{\infty}
e^{i(n-m)\phi_{lj}}\,H_{n-m}(\varkappa r_{lj})
\left(\begin{matrix} a_{ln}(K) \\ b_{ln}(K)
\end{matrix}\right)=
\left(\begin{matrix} p_{jm}(K) \\ q_{jm}(K)
\end{matrix}\right).
\label{MainSyst}
\end{equation}

Here $\varkappa=\sqrt{\omega^2-K^2}$, $r_{lj}$, $\phi_{lj}$ are
the polar coordinates of two-dimensional vector
$\mathbf{r}_{lj}=\{x_j-x_l,\,y_j-y_l\}$, $H_n(r)$ is the Hankel
function of the first kind, and $S_{jm}(\omega,K)$ is the
scattering matrix for the $j$-th waveguide. The scattering by a
cylindrical waveguide doesn't mix harmonics with different
longitudinal wave vectors $K$ and with different angular momenta
$m$, but harmonics of TE- and TM-types mix. The formulae to
calculate the scattering matrix $S_{jm}(\omega,K)$ are cited in
Appendix.

System (\ref{MainSyst}) allows to find partial amplitudes of waves
scattered by all the waveguides of the array. The spatial
distribution of field can be calculated by formulae

\begin{equation}
\begin{array}{l} \displaystyle
\mathbf{E}(t,\mathbf{r})=\mathbf{E}_{inc}(t,\mathbf{r})+
\sum\limits_{j=1}^N \mathbf{E}_{sca}(t,\mathbf{r}),
\medskip \\ \displaystyle
\mathbf{H}(t,\mathbf{r})=\mathbf{H}_{inc}(t,\mathbf{r})+
\sum\limits_{j=1}^N \mathbf{H}_{sca}(t,\mathbf{r}).
\end{array}
\label{EHtot}
\end{equation}

The exact system (\ref{MainSyst}) consists of infinite number of
equations, containing infinite number of variables. The number of
equations and variables can be limited, choosing some maximal
absolute value of angular momentum $m_{\max}$ and taking into
account only equations and partial amplitudes with $m$ lying in
interval $-m_{\max}\leq m\leq m_{\max}$. The spatial distribution
of field can be calculated with required accuracy choosing enough
great $m_{\max}$. It was demonstrated by the direct numerical
simulation, that $m_{\max}=2$ is enough for qualitative
description of optical excitation behaviour.

%%===============================================================

\section{Band structure calculation.}

Consider the infinite periodic plane array of identical
cylindrical waveguides. The array is situated in $xz$-plane, and
waveguides are directed along $z$-axis. The distance between two
adjacent waveguides is $a$. Below we discuss the eigenmodes of
this array and describe the method of band structure calculation.

The system of equations for eigenmodes has the left-hand side
coinciding with that in system (\ref{MainSyst}), and its
right-hand side is zero.

\begin{equation}
\Bigl(S_{jm}(\omega,K)\Bigr)^{-1} \left(\begin{matrix} a_{jm}(K)
\\ b_{jm}(K)
\end{matrix}\right)-
\underset{(l\neq j)}{\sum\limits_{l=-\infty}^{+\infty}}
\sum\limits_n e^{i(n-m)\phi_{lj}}\,H_{n-m}(\varkappa r_{lj})
\left(\begin{matrix} a_{ln}(K) \\ b_{ln}(K)
\end{matrix}\right)=0.
\label{EigModesSyst}
\end{equation}

Since the array is planar, $r_{lj}=a|l-j|$, $\phi_{lj}=0$ for
$j>l$ and $\phi_{lj}=\pi$ for $j<l$.

The eigenmodes of an infinite periodical array take the form of
Bloch waves characterized by the transversal quasi-wave vector $k$
($-\pi/a<k\leq \pi/a$):

\begin{equation}
a_{jm}(K)=a_m(k,K)\,e^{ikaj}, \qquad b_{jm}(K)=b_m(k,K)\,e^{ikaj}.
\label{Bloch}
\end{equation}

Substituting these expressions to (\ref{EigModesSyst}), a system
of equations for $a_m(k,K)$, $b_m(k,K)$ is obtained:

\begin{equation}
\sum\limits_n U_{mn}(\omega,k,K) \left(\begin{matrix} a_n(k,K) \\
b_n(k,K) \end{matrix}\right)=0, \label{EigModesSyst01}
\end{equation}

where

\begin{equation}
U_{mn}(\omega,k,K)=\Bigl(S_m(\omega,K)\Bigr)^{-1}\,
\delta_{mn}-\sum\limits_{j=1}^{+\infty}\,
\Bigl(e^{-ikaj}+(-1)^{n-m}\,e^{ikaj}\Bigr)\,H_{n-m}(\varkappa\,aj),\left(\begin{matrix}
1 && 0 \\ 0 && 1
\end{matrix}\right).
\label{U}
\end{equation}

The angular momentum takes values $-m_{\max}\leq m\leq m_{\max}$,
so (\ref{EigModesSyst01}) is a homogeneous linear system of
$4m_{\max}+2$ equations with the same number of variables. If this
system is represented in matrix form, its matrix $U(\omega,k,K)$
is composed of $(2m_{\max}+1)^2$ matrices $U_{mn}(\omega,k,K)$.
The system (\ref{EigModesSyst01}) possesses a nontrivial solution
when the matrix $U(\omega,k,K)$ is singular:
$\det\,U(\omega,k,K)=0$.

\begin{figure}[htbp] \centering
\includegraphics[width=0.8\textwidth]{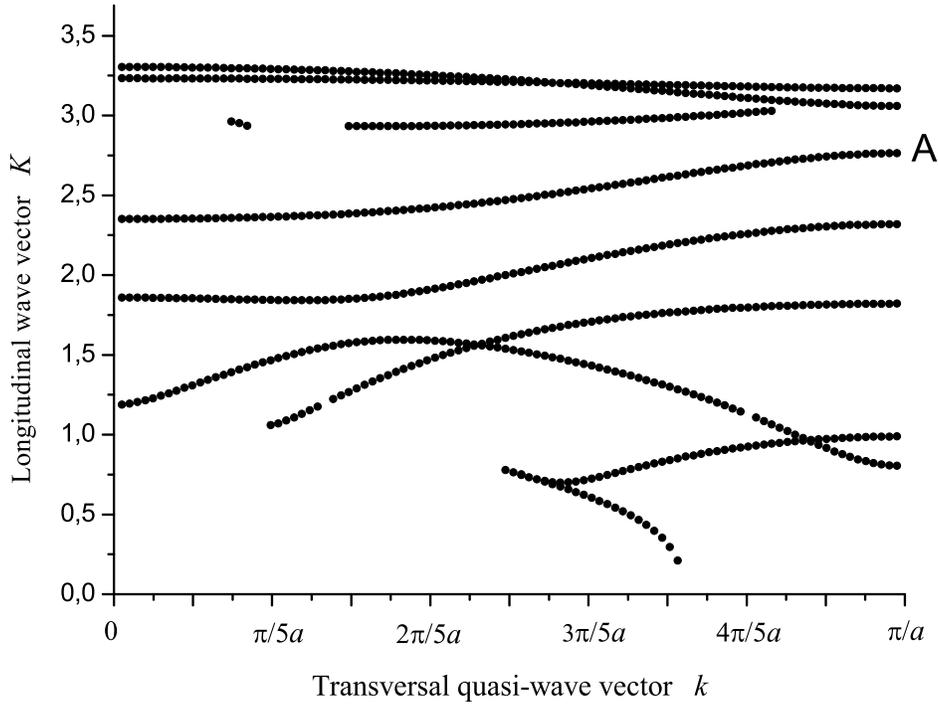}
\caption{Dispersion curves $K(k)$.} \label{FIGDispLaw}
\end{figure}

Using the technic described above, we calculated the band
structure for an array of cylindrical rods of unit radii ($R=1$)
made of GaAs (refractive index $n_r=3.5$). The rods are situated
next to each other, so the period of the array is $a=2R=2$. We
calculated the dependence of longitudinal wave vector $K$ on
transverse quasi-wave vector $k$ for a fixed frequency
$\omega=0.7\pi/a$ (below we will use the term ``dispersion law''
for the dependence $K(k)$). The approximation $m_{\max}=2$ was
used. It was shown by the direct numerical simulation, that for
the chosen $m_{\max}$ the dispersion law $K(k)$ can be found
accurate within 2\%. The dispersion curves $K(k)$ are presented at
Fig. \ref{FIGDispLaw}.

In Fig. \ref{FIGDispLaw} several dispersion curves corresponding
to several different bands are illustrated. Below we consider one
of the bands, that is noted by letter ``A''. This band is
convenient for further investigation, since it doesn't overlap
with other bands.

Since the parameters $K$ and $k$ of eigenmodes are connected by
the dispersion laws, one of arguments in notations $a_m(k,K)$,
$b_m(k,K)$ for partial amplitudes is unnecessary, so below the
partial amplitudes are denoted $a_m(K)$, $b_m(K)$.

%%===============================================================

\section{Bloch oscillation prediction on basis of dispersion law.}

Below we consider the Gaussian beam propagating in the array of
waveguides. The partial amplitudes describing this excitation are
represented by formulae

\begin{equation}
\begin{array}{c} \displaystyle
a_m(K)=a_m\,e^{-\tau^2\,(K-K_0)^2},
\medskip \\ \displaystyle
b_m(K)=b_m\,e^{-\tau^2\,(K-K_0)^2}.
\end{array}
\label{Beam001}
\end{equation}

In this case the field distribution takes the form

\begin{equation}
\begin{array}{c} \displaystyle
\mathbf{E}(t,\mathbf{r})= e^{-i\omega t}\mathbf{u}(\mathbf{r})\,
\exp\left\{-\frac{(x/v-z)^2}{4\tau^2}+i\,k_0\,x
+i\,K_0\,z\right\},
\medskip \\ \displaystyle
\mathbf{H}(t,\mathbf{r})= e^{-i\omega t}\mathbf{v}(\mathbf{r})\,
\exp\left\{-\frac{(x/v-z)^2}{4\tau^2}+i\,k_0\,x
+i\,K_0\,z\right\}.
\end{array}
\label{Beam002}
\end{equation}

Here $\mathbf{u}(\mathbf{r})$, $\mathbf{v}(\mathbf{r})$ are the
functions periodically depending on $x$, $k_0$ is connected with
$K_0$ by the dispersion law, $K_0=K(k_0)$, and $v=dK/dk(k_0)$. The
formulae (\ref{Beam002}) are correct for enough large values of
$\tau$.

It follows from Eqs (\ref{Beam002}), that in the periodical array
of identical waveguides the optical excitation propagates along
the straight line $x(z)=vz$, and the direction of propagation is
defined by the dispersion law $K(k)$.

But the situation changes dramatically, if the optical
characteristics of waveguides (such as thickness or refractive
index) gradually vary across the array.

One can mentally divide the array to sections much wider than the
optical beam, but enough narrow for one could assume the
waveguides into a section to be identical. One can attribute a
local dispersion law $K(k)$ to every section. Therefore, the
direction of the beam propagation should be different in different
sections, and the propagation way of optical excitation should be
curved. A certain form of the propagation way can be predicted by
calculating the dispersion law for arrays with different
refractive indices of the waveguides.

For example, consider an array of $N=100$ cylindrical rods of
unite radii. The refractive index of a rod in the middle of the
array ($j=50$) is $n_r^{50}=3.5$, and the difference between the
refractive indices of two adjacent waveguides is
$n_r^j-n_r^{j+1}=0.01$, i. e.

\begin{equation}
n_r^j=3.5-0.01\,(j-50) \label{nrj}
\end{equation}

The section at the middle of the array is similar to the array
considered in the previous section. So, we choose the parameters
of optical beam according to the dispersion curve represented in
Fig. \ref{FIGDispLaw} and marked by letter ``A''. The band
corresponding to that dispersion curve lies in the range
$2.35<K<2.78$, so we choose $K_0=2.565$ exactly at the middle of
the band. The frequency of the excitation $\omega=0.35\pi$.

\begin{figure}[htbp] \centering
\includegraphics[width=0.8\textwidth]{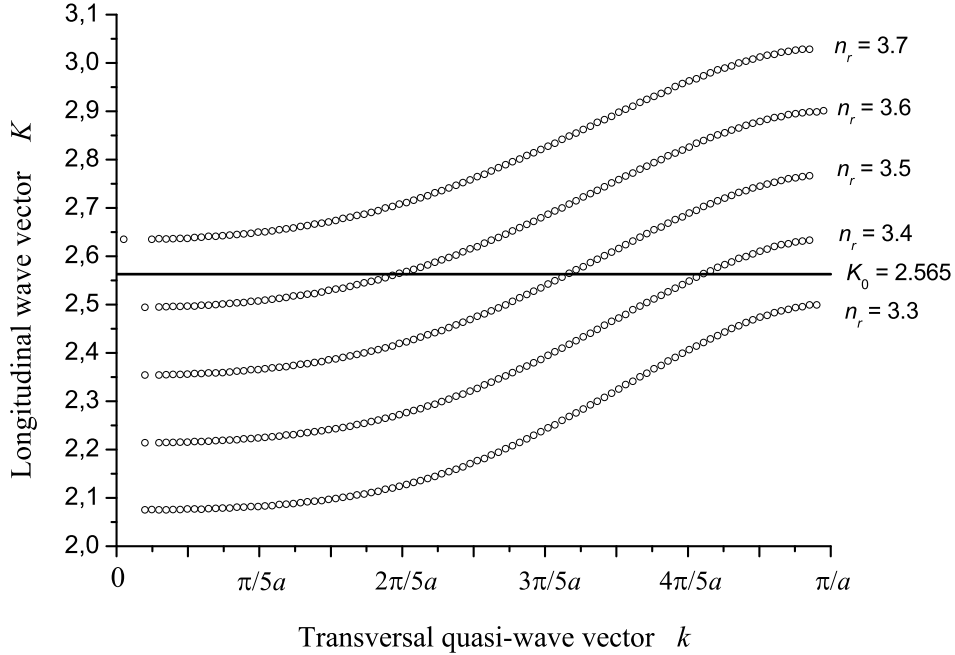}
\caption{Dispersion laws for different refractive indices. The
straight line corresponds to $K=2.565$.} \label{FIGDispLawsDiffNR}
\end{figure}

If the refractive index changes, the dispersion curve ``A''
shifts, as it is shown in Fig. \ref{FIGDispLawsDiffNR}. We have
found that the longitudinal wave-vector $K_0$ lies into the band
``A'' if the refractive index varies in the interval
$3.35<n_r<3.65$. Therefore, the optical excitation can propagate
in a part of the array where the refractive indices of waveguides
belong to the mentioned interval, i.e. between the 35-th and 65-th
waveguides.

So, we have predicted the amplitude of Bloch oscillation. But one
can also predict the period of Bloch oscillation and the way of
optical beam propagation. For this purpose, the value $v=dK/dk$
for $K_0$ for different values of refractive index $n_r$ should be
calculated. For the obtained dependence the notation $v(n_r)$ will
be used. The way $x(z)$ of optical beam propagation is determined
by the differential equation

\begin{equation}
\frac{dx}{dz}=v\Bigl(n_r(x)\Bigr), \label{Beam003}
\end{equation}

where the function $n_r(x)$ is obtained by the interpolation of
dependence of the waveguide refractive index $n_r^j$ on the
waveguide number $j$:

\begin{equation}
n_r(x)=3.5-0.01\left(\frac{x}{a}-50\right). \label{Beam004}
\end{equation}

\begin{figure}[htbp] \centering
\includegraphics[width=0.8\textwidth]{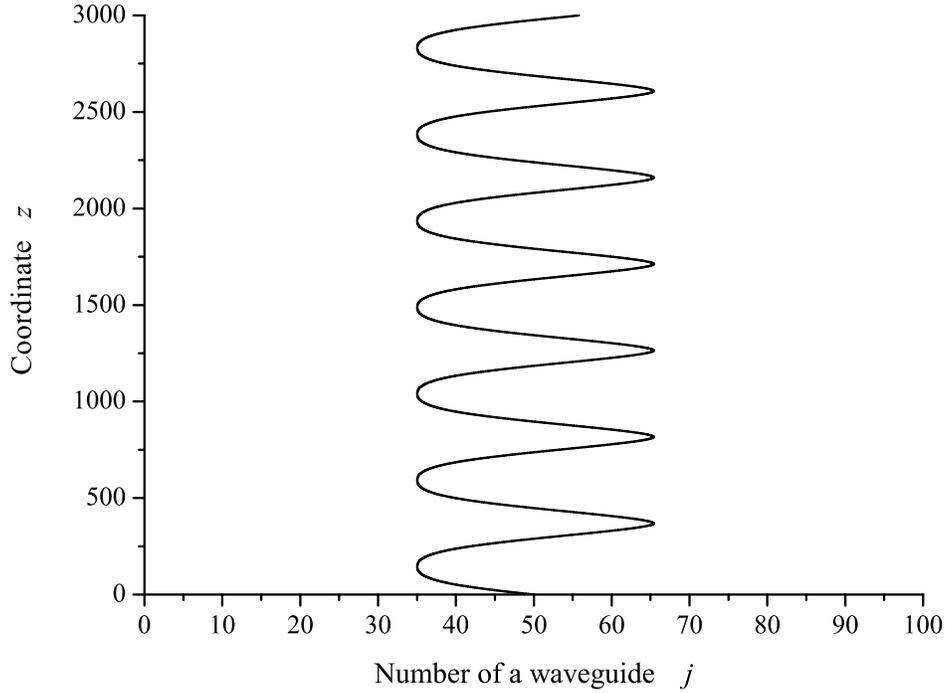}
\caption{Optical excitation propagation way obtained by the
analysis of dispersion law.} \label{FIGBlochOscOnDispLaw}
\end{figure}

Eq. (\ref{Beam003}) can be integrated numerically. The way of
optical beam propagation, obtained from this equation, has the
form of periodical oscillation, as represented in Fig.
\ref{FIGBlochOscOnDispLaw}. The period of the obtained oscillation
is $\Delta z \approx 220a=440$ (remind that $a=2$).

%%===============================================================

\section{Direct calculation of Bloch oscillation and breathing mode.}

In this section we represent the results of direct calculation of
Gaussian beam propagation, based on numerical solution of Eq.
\ref{MainSyst}. We consider the same array as in the previous
section. The array is illuminated by an incident wave that is
defined by partial amplitudes

\begin{equation}
\begin{array}{c} \displaystyle
p_{jm}(K)=p_m\,e^{-\tau^2\,(K-K_0)^2}\,
\exp\left\{-\frac{a^2\,(j-j_0)^2}{4\sigma^2}
+ik_0a\,(j-j_0)\right\},
\medskip \\ \displaystyle
q_{jm}(K)=q_m\,e^{-\tau^2\,(K-K_0)^2}\,
\exp\left\{-\frac{a^2\,(j-j_0)^2}{4\sigma^2}
+ik_0a\,(j-j_0)\right\}.
\end{array}
\label{IncWave}
\end{equation}

This incident wave illuminates the finite area of the array. The
parameters $\sigma$ and $\tau$ define the width of the illuminated
area along $x$-axis and $z$-axis correspondingly. We take
$j_0=50$, i. e. the incident wave illuminates the middle of the
array. The parameters $K_0$ and $k_0$ are connected by the
dispersion law marked by letter ``A'' in Fig. \ref{FIGDispLaw}. We
take $K_0=2.565$, exactly at the middle of the band ``A''
($2.35<K<2.78$), and the corresponding $k_0=0.989$. The parameter
$\tau$ is chosen so that the peak of the function
$e^{-\tau^2\,(K-K_0)^2}$ fits into the band ``A'': $\tau=6/\Delta
K \approx 14$, where $\Delta K$ is the width of the band ``A''.
The parameter $\sigma=v(k_0)\,\tau$, where $v(k_0)$ is defined by
the dispersion law: $v(k_0)=dK/dk(k_0)$. Here $v(k_0)\approx 0.5$,
so $\sigma=7$.

The parameters $p_m$, $q_m$ are chosen so that the incident wave
excites the eigenmodes of the array effectively. For our
calculation we chose $q_0=1$ and all the other $p_m$, $q_m$ are
zeros. The computation is performed for the approximation
$m_{\max}=2$.

\begin{figure}[htbp] \centering
\includegraphics[width=\textwidth]{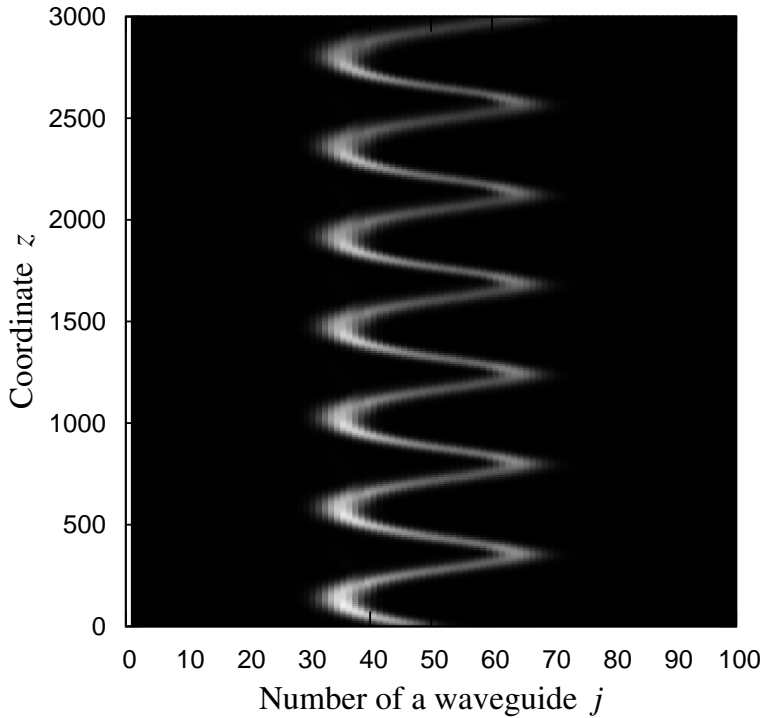}
\caption{Optical excitation propagation.} \label{FIGBlochOsc}
\end{figure}

The result of the computation is presented in Fig.
\ref{FIGBlochOsc}. As expected, the obtained way of optical beam
propagation has a periodical form. It oscillates between the 35-th
and 65-th waveguides, and the period of oscillation is $\Delta
z=440$. It is remarkable that the form of oscillation obtained by
the numerical solution of Eq. (\ref{MainSyst}) coincides exactly
with that obtained by the dispersion law analysis.

Besides the Bloch oscillation, we consider the so-called breathing
mode \cite{Wang1,Morandotti,Dreisow1}. Such kind of optical
excitation arises when only one waveguide of the array is
illuminated by the incident wave. The characteristic feature of
breathing mode is the periodical spreading and focusing behaviour.

\begin{figure}[htbp] \centering
\includegraphics[width=1\textwidth]{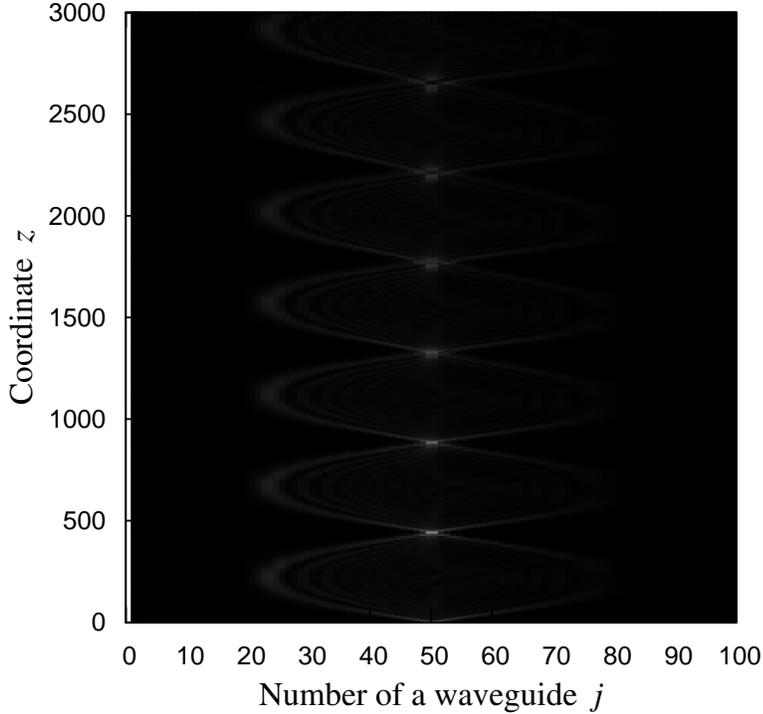}
\caption{Breathing mode.} \label{FIGBreathingModes}
\end{figure}

We assume that the incident wave illuminates a short section
around $z=0$ of the 50-th waveguide situated at the middle of the
array. To simulate this situation, we take the partial amplitudes
of the incident wave as follows: $p_{jm}(K)=0$, $\underset{m\neq
0}{q_{jm}}(K)=0$, $q_{j0}(K)=1$. The result of direct numerical
calculation is represented in Fig. \ref{FIGBreathingModes}.

%%===============================================================

\section{Bloch-Zener oscillation.}

The optical Bloch-Zener oscillation in an array of optical
waveguides can take place if the band structure consists of
several bands separated by gaps. If the refractive index of
waveguides gradually varies across the array, the band structures
of two different sections of the array are shifted relative to
each other. Therefore, the lower band of one section can overlap
the upper band of another section. So, the optical beam can
partially tunnel from one section to another. This leads to that
the optical beam divides into two beams propagating along two
different oscillating ways.

To demonstrate this effect, it is convenient to consider an array
with the band structure containing two bands separated with a
narrow gap. We take the array of waveguides of two types situated
by turns. The refraction indices of waveguides of the first and
the second type are $n_{r1}=3.5$ and $n_{r2}=3.55$ respectively.
The band structure of this array contains two bands $2.38<K<2.48$
and $2.54<K<2.8$ (we suppose the frequency $\omega=0.35\pi$, as in
previous sections).

We introduce a small variation of refraction indices of
waveguides:

\begin{equation}
\begin{array}{c} \displaystyle
n_r^j=3.5-0.005\,(j-50) ~~ \text{ for odd } ~ j,
\medskip \\ \displaystyle
n_r^j=3.55-0.005\,(j-50) ~~ \text{ for even } ~ j.
\end{array}
\label{NREvenOdd}
\end{equation}

The array is illuminated by the incident wave defined by the
partial amplitudes $p_{jm}(K)$, $q_{jm}(K)$, that are given by
formulae (\ref{IncWave}). The parameters $K_0$, $k_0$, $\tau$,
$\sigma$ entering to these formulae are chosen according to the
principle similar to that described in the previous section. The
parameter $K_0=2.67$ is taken exactly at the middle of the band
$2.54<K<2.8$, the parameter $k_0=0.44$ is connected to $K_0$ by
the dispersion law $K(k)$, the value of $\tau=23$ is taken so that
the peak of the function $e^{-\tau^2\,(K-K_0)^2}$ entirely fits
into the band $2.54<K<2.8$, and $\sigma=dK/dk(k_0)\,\tau=10$.

\begin{figure}[htbp] \centering
\includegraphics[width=1\textwidth]{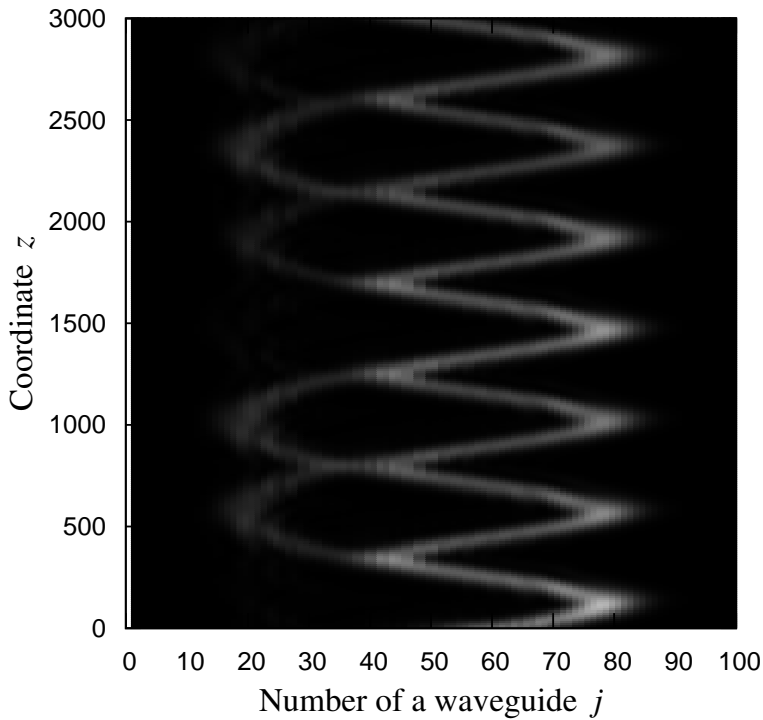}
\caption{Bloch-Zener oscillation.} \label{FIGBlochZenerOsc}
\end{figure}

The result of the calculation is represented on Fig.
\ref{FIGBlochZenerOsc}.

%================================================================

\section{Conclusion.}

In this paper three phenomena are considered --- Bloch
oscillation, Bloch-Zener oscillation and breathing modes in planar
arrays of optical waveguides with gradually varying refractive
index. We suggest a new method to investigate this subject, based
on the multiple scattering formalism. This method has several
advantages over the traditional method based on Eq. (\ref{Tradi}).
The MSF allows to find the spatial distribution of field with any
required accuracy, while the traditional method gives only the
intensity of optical excitation. Besides, the input data for MSF
are the geometrical properties of the array and refractive indices
of waveguides, while the traditional method requires some data
that should be obtained experimentally, such as the longitudinal
wave vectors of eigenmodes of waveguides and coupling constants.

The MSF represented in this paper is convenient only for the
waveguides of cylindrical form, because in this case the
scattering matrix can be calculated easily. However, this method
can be applied for the waveguides of another shape, but in this
case it would be more difficult to calculate the scattering
matrix. Besides, the scattering by noncylindrical waveguides would
mix the harmonics with different angular momenta. So, if the shape
of the waveguides is enough complicated, one should take into
account the harmonics with enough high angular momenta, and the
calculation would be difficult. At the same time, for the
cylindrical waveguides it is enough to take into account the
harmonics with $|m|\leq 2$, as it is shown in this work.

The considered phenomena may be useful for different optical
applications, such as steering, splitting, focusing and defocusing
of light. The method represented in this work allows to produce
the numerical simulation without need of experimental
investigation of components of optical devices.

%%===============================================================

%%===============================================================

\section*{APPENDIX.}

\subsection*{1. Formulae for functions $\mathbf{M}^{(1)}_{\omega Km}(r)$
and $\mathbf{N}^{(1)}_{\omega Km}(r)$:}

\begin{equation}
\begin{array}{c} \displaystyle
\mathbf{M}^{(1)}_{\omega Km}(r)=

\mathbf{e}_r\,\frac{iK}{2\varkappa}\, \Bigl(J_{m-1}(\varkappa
r)-J_{m+1}(\varkappa r)\Bigr)+
\medskip \\ \displaystyle
+\mathbf{e}_\phi\,\frac{-K}{2\varkappa}\, \Bigl(J_{m-1}(\varkappa
r)+J_{m+1}(\varkappa r)\Bigr)%+
%%\medskip \\ \displaystyle
+\mathbf{e}_z iJ_m(\varkappa r),
\end{array}
\label{AppM1}
\end{equation}

\bigskip

\begin{equation}
\begin{array}{c} \displaystyle
\mathbf{N}^{(1)}_{\omega Km}(r)=
\mathbf{e}_r\,\frac{\omega}{2\varkappa}\, \Bigl(J_{m-1}(\varkappa
r)+J_{m+1}(\varkappa r)\Bigr)+
\medskip \\ \displaystyle
+\mathbf{e}_\phi\,\frac{i\omega}{2\varkappa}\,
\Bigl(J_{m-1}(\varkappa r)-J_{m+1}(\varkappa r)\Bigr).
\end{array}
\label{AppN1}
\end{equation}

Here $\varkappa=\sqrt{\omega^2-K^2}$, $J_m(x)$ is Bessel function.

\subsection*{2. Formulae for functions $\mathbf{M}^{(2)}_{\omega Km}(r)$
and $\mathbf{N}^{(2)}_{\omega Km}(r)$:}

\begin{equation}
\begin{array}{c} \displaystyle
\mathbf{M}^{(2)}_{\omega Km}(r)=
\mathbf{e}_r\,\frac{iK}{2\varkappa}\, \Bigl(H_{m-1}(\varkappa
r)-H_{m+1}(\varkappa r)\Bigr)+
\medskip \\ \displaystyle
+\mathbf{e}_\phi\,\frac{-K}{2\varkappa}\, \Bigl(H_{m-1}(\varkappa
r)+H_{m+1}(\varkappa r)\Bigr)%+
%%\medskip \\ \displaystyle
+\mathbf{e}_z iH_m(\varkappa r),
\end{array}
\label{AppM2}
\end{equation}

\bigskip

\begin{equation}
\begin{array}{c} \displaystyle
\mathbf{N}^{(2)}_{\omega Km}(r)=
\mathbf{e}_r\,\frac{\omega}{2\varkappa}\, \Bigl(H_{m-1}(\varkappa
r)+H_{m+1}(\varkappa r)\Bigr)+
\medskip \\ \displaystyle
+\mathbf{e}_\phi\,\frac{i\omega}{2\varkappa}\,
\Bigl(H_{m-1}(\varkappa r)-H_{m+1}(\varkappa r)\Bigr).
\end{array}
\label{AppN2}
\end{equation}

Here $H_m(x)$ is Hankel function of the first kind.

\subsection*{3. Formulae for scattering matrix:}

Consider an infinite dielectric rod situated along the $z$-axis.
The radius of the rod is $R$, and its refractive index $n_r$. It
is illuminated by a monochromatic wave of frequency $\omega$ with
certain longitudinal wave vector $K$ and angular momentum $m$.
This wave is defined by two partial amplitudes $p_m(K)$, $q_m(K)$.
The scattered wave possesses the same frequency $\omega$,
longitudinal wave vector $K$ and angular momentum $m$. It is
defined by partial amplitudes $a_m(K)$, $b_m(K)$.

Partial amplitudes of incident and scattered waves are connected
by the scattering matrix $S_m(\omega,K)$:

\begin{equation}
\left(\begin{matrix} a_m(K) \\ b_m(K)
\end{matrix}\right)=S_m(\omega,K) \, \left(\begin{matrix} p_m(K)
\\ q_m(K) \end{matrix}\right). \label{AppS1}
\end{equation}

\bigskip

To formulate the expression for matrix $S_m(\omega,K)$, we
introduced some notations:

\begin{equation}
\begin{array}{c} \displaystyle
\varkappa=\sqrt{\omega^2-K^2}, ~~~~~ \xi=\varkappa R, ~~~~~
\alpha=K/2\varkappa, ~~~~~ \beta=\omega/2\varkappa,
\\ \displaystyle
\varkappa_i=\sqrt{n_r^2\,\omega^2-K^2}, ~~~~~ \xi_i=\varkappa_i R,
~~~~~ \alpha_i=K/2\varkappa_i, ~~~~~~~
\beta_i=n_r\omega/2\varkappa_i,
\end{array}
\label{AppNot0}
\end{equation}

\bigskip

\begin{equation}
\begin{array}{c} \displaystyle
w_-=J_{m-1}(\xi)-J_{m+1}(\xi), ~~~~~~~ w_0=J_m(\xi), ~~~~~~~
w_+=J_{m-1}(\xi)+J_{m+1}(\xi),
\\ \displaystyle
u_-=H_{m-1}(\xi)-H_{m+1}(\xi), ~~~~~~~ u_0=H_m(\xi), ~~~~~~~
u_+=H_{m-1}(\xi)+H_{m+1}(\xi),
\\ \displaystyle
v_-=J_{m-1}(\xi_i)-J_{m+1}(\xi_i), ~~~~~~~ v_0=J_m(\xi_i), ~~~~~~~
v_+=J_{m-1}(\xi_i)+J_{m+1}(\xi_i).
\end{array}
\label{AppNot1}
\end{equation}

\bigskip

\begin{equation}
\begin{array}{c}
M_{11}=\left(\begin{matrix} -i\alpha\,u_- & \beta\,u_+ \\
\alpha\,u_+ & i\beta\,u_-\end{matrix}\right), \qquad
M_{21}=\left(\begin{matrix} -u_0 & 0 \\ 0 & -u_0
\end{matrix}\right),
\medskip \\ \displaystyle
M_{12}=\left(\begin{matrix} in_r^2\alpha_i\,v_- &
-n_r^2\beta_i\,v_+
\\ -\alpha_i\,v_+ & -i\beta_i\,v_-\end{matrix}\right), \qquad
M_{22}=\left(\begin{matrix} v_0 & 0 \\ 0 & n_rv_0
\end{matrix}\right),
\medskip \\ \displaystyle
N_1=\left(\begin{matrix} i\alpha\,w_- & -\beta\,w_+ \\
-\alpha\,w_+ & -i\beta\,w_-\end{matrix}\right), \qquad
N_2=\left(\begin{matrix} w_0 & 0 \\ 0 & w_0
\end{matrix}\right).
\end{array}
\label{AppNot2}
\end{equation}

\bigskip

Using the introduced notations, we write down the expression for
matrix $S_m(\omega,K)$:

\begin{equation}
S_m(\omega,K)=\Bigl(M_{12}^{-1}\,M_{11}-M_{22}^{-1}\,M_{21}\Bigr)^{-1}\,
\Bigl(M_{12}^{-1}\,N_1-M_{22}^{-1}\,N_2\Bigr). \label{AppS2}
\end{equation}

\end{document}